\documentclass{appolb}
\usepackage{epsfig}
\newcommand{\as}{\alpha_{\rm s}}

\def\Lm{\mathrm{L}_m}

\def\Lx{\mathrm{L}_x}
\def\Ly{\mathrm{L}_y}

\begin{document}
\title{NNLO Virtual corrections to W$^+$ W$^-$ production at the LHC%
\thanks{Presented at ``Matter To The Deepest:
Recent Developments In Physics
of Fundamental Interactions'', Ustron, 5-11 September 2007, Poland}%
}
\author{Grigorios Chachamis
\address{Institut f\"ur Theoretische Physik und Astrophysik,
Universit\"at W\"urzburg\\
Am Hubland, D-97074 W\"urzburg, Germany}
%\and
%the Name(s) of other Author(s)
%\address{and their affiliation}
}
\maketitle
\begin{abstract}
We report on a recent calculation of the two-loop virtual QCD
corrections to the W boson pair production in the 
quark-anti-quark-annihilation channel in the limit where all 
kinematical invariants are large compared to the mass of the W boson.
Our result is exact up to terms suppressed by powers of the 
W boson mass. The infrared pole structure is in agreement with the
prediction of Catani's general formalism for the singularities of
two-loop QCD amplitudes.

\end{abstract}

  
\section{Introduction}

W pair production via quark-anti-quark-annihilation
is a very important process at the Large Hadron Collider (LHC) for
two main reasons. Firstly, it can serve as a signal process 
in the search for new Physics since it can be used
to measure the vector boson trilinear couplings as predicted by the
Standard Model (SM). Deviations from these predictions may come from
decays of new heavy particles into SM vector boson pairs or anomalous
couplings and would signal new Physics~\cite{tevatron}.
Secondly, $q {\bar q} \rightarrow W^+ W^-$ is the dominant irreducible
background to the promising Higgs discovery channel 
$p p \rightarrow H \rightarrow W^* W^* \rightarrow l {\bar \nu} {\bar l}' \nu'$
in the mass range M$_{\mathrm Higgs}$ 
between 140 and 180 GeV~\cite{dittmardreiner}.
In this paper, we describe the computation of the two-loop virtual QCD
corrections to the W boson pair production in the 
quark-anti-quark-annihilation channel. The full result will appear 
in~\cite{ccd}.

Due to its importance, the study of W pair production in hadronic 
collisions has
attracted a lot of attention in the literature. The Born cross section was
computed almost 30 year ago~\cite{brown}, whereas the 
next-to-leading order (NLO)
QCD corrections to the tree-level were computed 
in Refs.~\cite{ohn,fri,dixon1,dixon2,campbell}
and were proven to be large. 
They enhance the tree-level by
almost 70\% which falls to a (still) large 30\% after 
imposing a jet veto. Another process that
adds to the $p p \rightarrow W^+ W^-$ background
is the W pair production
in the loop induced gluon fusion channel, $g g \rightarrow W^+ W^-$. This 
contributes at $\mathcal{O}(\alpha_s^2)$ relative to the 
quark-anti-quark-annihilation channel but is 
nevertheless enhanced due to the large gluon flux
at the LHC. The corrections from gluon fusion increase the W pair background
estimate by almost  30\% after certain experimental Higgs search cuts 
are imposed~\cite{kauer1,kauer2}.

Given that the NLO QCD corrections to the background are large
and also that the cross section for the process
$H \rightarrow W W \rightarrow l {\bar \nu} {\bar l}' \nu'$ (signal
process for the Higgs discovery) is known at 
NNLO~\cite{Anastasiou:2007mz}, the NNLO corrections to 
$q {\bar q} \rightarrow W^+ W^-$ need to be computed. 
This will allow a theoretical estimate for W production from
$q {\bar q}$-annihilation with accuracy better than
10\%,  as well as having both the signal and the background process
calculated at the same order of the perturbative expansion (NNLO).
In this paper, we address the task
of computing the NNLO two-loop virtual part, more precisely
the interference of the two-loop with the Born amplitude.
We work in the limit of fixed scattering angle and high energy, where all
kinematical invariants are large compared to the mass $m$ of the W.

\section{The calculation}
\label{sec:compdetails}

Our methodology for obtaining
the massive amplitude (massless fermion-boson scattering was studied 
in Ref.~\cite{Anastasiou:2002zn}) 
is very similar to the one followed 
in Refs.~\cite{qqTT,ggTT,Czakon:2004wm}, an evolution actually,
of the methods employed in Refs.~\cite{Czakon:2006pa,Actis:2007gi}.
The amplitude is reduced to a form that 
contains only a small number of integrals (master integrals)
with  the help of the Laporta algorithm~\cite{Laporta:2001dd}.
In our calculation there are 71 master integrals.
Next comes the construction, in a fully automatic way,
of the Mellin-Barnes (MB) 
representations~\cite{Smirnov:1999gc,Tausk:1999vh}
of all the master integrals using
the {\tt MBrepresentation} package~\cite{MBrepresentation}. The
representations  are then
analytically continued in the dimension of space-time with the help of the
{\tt MB} package~\cite{Czakon:2005rk} revealing the full singularity
structure. An asymptotic 
expansion in the mass parameter is performed by
closing contours and the integrals are finally resummed,
either with the help of {\tt XSummer}~\cite{Moch:2005uc},
or the {\tt PSLQ} algorithm~\cite{pslq:1992}.

\begin{figure}[ht]
  \begin{center}
    \epsfig{file=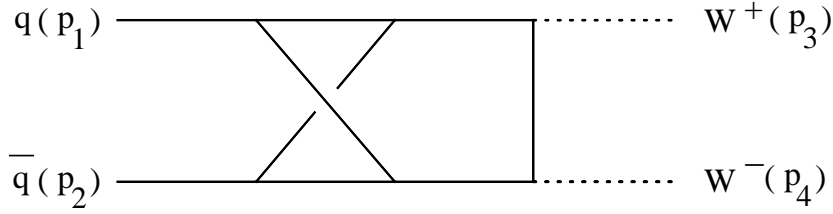,width=11cm}
    \caption{\label{fig:masterPR0}
A non-planar two-loop box master integral with two massive legs.
$p_j$ are the external momenta, with $p_1 = p_2 = 0$ and 
$p_3 = p_4 = m^2$, $m$ the mass of the W.
    }
  \end{center}
\end{figure}

We will present here the result for a non-planar
two-loop box master integral with two massive legs 
(Fig.~\ref{fig:masterPR0}).
The Mellin-Barnes representation of this integral is 7-fold and reads:
\begin{eqnarray}
I_{NP} &=& - (-s)^{-3 - 2 \epsilon} 
\int_{- i \infty}^{i \infty} \prod_{j = 1}^{7} d z_j
\left(\frac{s}{m^2}\right)^{-z1} 
\left(\frac{t}{m^2}\right)^{-z2}
\left(\frac{u}{m^2}\right)^{-z3} 
\nonumber \\ && \times
\Gamma(-\epsilon)^2 
\Gamma(1 + z_1) 
\Gamma(z_2) 
\Gamma(z_3) 
\Gamma(-z_4)
\Gamma(-z_5) 
\nonumber \\ && \times
\Gamma(-z_1 - z_2 - z_3 + z_4 + z_5) 
\Gamma(-1 -\epsilon - z_4 - z_6)
\Gamma(-z_6)
 \nonumber \\ && \times
\Gamma(-2 - 2\epsilon - z_1 - z_7) 
\Gamma(-z_5 - z_7) 
\Gamma(-z_6 - z_7)
\nonumber \\ && \times
\Gamma(-2 - 2\epsilon + z_2 - z_4 - z_5 - z_6 - z_7) 
\Gamma(1 - z_2 + z_5 + z_7) 
\nonumber \\ && \times
\Gamma(-2 - 2\epsilon + z_3 - z_4 - z_5 - z_6 - z_7) 
\Gamma(1 - z_3 + z_5 + z_7)
\nonumber \\ && \times
\Gamma(3 + 2 \epsilon + z_1 + z_6 + z_7)
\Gamma(1 + z_4 + z_6 + z_7)
\nonumber \\ && \times
\left( 
\Gamma(-1 - 3 \epsilon) 
\Gamma(-2 \epsilon) 
\Gamma(-1 - 2 \epsilon - z_4 - z_6)^2
\right. \nonumber \\ && \times \left.
\Gamma(-z_5 - z_6 - z_7)
\right)^{-1}\, .
\end{eqnarray}
This specific master integral is needed up to order
${\mathcal{O}}((m^2/s)^{-1})$.
After the analytical continuation in the spacetime dimension and 
the expansion in the mass one is able to perform the resummation 
which yields:
\begin{eqnarray}
I_{NP} &=& 
+ \frac{1}{2 m_s \epsilon^4} 
+ \frac{1}{m_s \epsilon^3} 
\left\{
- 2 \Lm 
+ \Lx 
+ \Ly
\right\}
\nonumber \\ &&
+ \frac{1}{m_s \epsilon^2} 
\left\{
  4 \Lm^2 
- 4 \Lx \Lm 
- 4 \Ly \Lm 
+ \Lx^2
+ \Ly^2
+ 2 \Lx \Ly
\right. \nonumber \\ &&  \left.
- \frac{89 \pi ^2}{12}
\right\}
%\nonumber \\ &&
+ \frac{1}{m_s \epsilon} 
\left\{ 
- \frac{16 \Lm^3}{3}
+ 8 \Lx \Lm^2
+ 8 \Ly \Lm^2
- 4 \Lx^2 \Lm
\right. \nonumber \\ &&  \left.
- 4 \Ly^2 \Lm
- 8 \Lx \Ly \Lm
+ \frac{89 \pi ^2 \Lm}{3}
+ \frac{2 \Lx^3}{3}
+ \frac{2 \Ly^3}{3}
+ 2 \Lx \Ly^2
\right. \nonumber \\ &&  \left.
+ 2 \Lx^2 \Ly
- \frac{40 \zeta_3}{3}
- \frac{89 \Lx \pi ^2}{6}
- \frac{89 \Ly \pi ^2}{6}
\right\}
+ \frac{1}{m_s}
\left\{
\frac{16 \Lm^4}{3}
-\frac{32 \Lx \Lm^3}{3}
\right. \nonumber \\ &&  \left.
-\frac{32 \Ly  \Lm^3}{3}
+ 8 \Lx^2 \Lm^2
+ 8 \Ly^2   \Lm^2
+ 16 \Lx \Ly \Lm^2
- \frac{178 \pi ^2   \Lm^2}{3}
\right. \nonumber \\ &&  \left.
- \frac{8 \Lx^3 \Lm}{3}
- \frac{8 \Ly^3 \Lm}{3}
- 8 \Lx \Ly^2 \Lm
- 8 \Lx^2 \Ly  \Lm
+ \frac{160 \zeta_3 \Lm}{3}
\right. \nonumber \\ &&  \left.
+ \frac{178 \Lx \pi ^2   \Lm}{3}
+ \frac{178 \Ly \pi ^2 \Lm}{3}
+ \frac{\Lx^4}{3}
+ \frac{\Ly^4}{3}
+ \frac{4 \Lx \Ly^3}{3}
+ 2 \Lx^2 \Ly^2
\right. \nonumber \\ &&  \left.
+ \frac{4 \Lx^3 \Ly}{3}
- \frac{80 \Lx   \zeta_3}{3}
- \frac{80 \Ly \zeta_3}{3}
+ \frac{1111 \pi ^4}{720}
- \frac{89 \Lx^2 \pi ^2}{6}
\right. \nonumber \\ &&  \left.
- \frac{89 \Ly^2 \pi ^2}{6}
- \frac{89 \Lx \Ly \pi ^2}{3}
\right\}\, .
\end{eqnarray}
We have used above the compact notation $m_s = m^2/s$,
$\mathrm{L}_m = \log(m^2/s)$, $\mathrm{L}_x = \log(x)$ and
$\mathrm{L}_y = \log(1-x)$. $s = (p_1 + p_2)^2$ and
$t = (p_1 - p_3)^2$ are the usual Mandelstam variables and $x = t/s$.
One of the checks in our calculation was
to verify numerically the analytic results
for the masters.

The renormalization of the amplitude
involves only the renormalization of the strong coupling 
constant $\as$~\cite{van Ritbergen:1997va,Czakon:2004bu}.
A non-trivial check for our computation is that 
the infrared pole structure of our 
renormalized result is in agreement with the
prediction of Catani's general formalism for the singularities of the
two-loop QCD amplitudes as described in Ref.~\cite{catani}.

\section{Outlook}
We have calculated  the
two-loop virtual QCD corrections to the W boson pair production
in the quark-anti-quark-annihilation channel in the high energy limit.
The two-loop result, along with the square of the one-loop
$2 \to 2$ process have to be combined with the 
tree-level
$2 \to 4$ and the one-loop $2 \to 3$ processes
in order to obtain physical
cross sections. 
Combining all these contributions will enable
the analytic cancellation of the remaining
infrared divergences. 
Initial state singularities will have to be absorbed into parton 
distribution functions of the hadrons (protons)
in order to match with a precise parton evolution 
at NNLO~\cite{Moch:2004pa,Vogt:2004mw}.

{\bf{Acknowledgments:}}
This work was performed with support of the Sofja Kovalevskaja programme
of the Alexander von Humboldt Foundation.

\end{document}